\documentclass[aps,pre,twocolumn,groupedaddress,superscriptaddress,showpacs]{revtex4-1}

\usepackage{graphicx}
\usepackage{color}
\usepackage{amsmath}
\usepackage{amsfonts}
\usepackage{amssymb}
\usepackage{dcolumn}
\usepackage{eqnarray}
\usepackage{hyperref}
\hypersetup{colorlinks=true,urlcolor=blue,linkcolor=blue,citecolor=blue}
\begin{document}
\title{Complex networks from space-filling bearings}
\author{J.~J.~Kranz} 
\email{j.kranz@kit.edu}
\affiliation{Computational Physics for Engineering Materials, IfB, ETH
Zurich, Wolfgang-Pauli-Strasse 27, CH-8093 Zurich, Switzerland}
\affiliation{Theoretical Chemical Biology, Institute of Physical
Chemistry, Karlsruhe Institute of Technology, Kaiserstr. 12, D-76131
Karlsruhe, Germany }
\author{N.~A.~M.~Ara\'ujo}
\email{nmaraujo@fc.ul.pt}
\affiliation{Departamento de F\'{\i}sica, Faculdade de
Ci\^{e}ncias, Universidade de Lisboa, P-1749-016 Lisboa, Portugal}
\affiliation{Centro de F\'isica Te\'orica e Computacional, Universidade de Lisboa,
P-1749-003 Lisboa, Portugal}
\author{J.~S.~Andrade~Jr.}
\email{soares@fisica.ufc.br}
\affiliation{Departamento de F\'{\i}sica, Universidade Federal do Cear\'a, 60451-970 Fortaleza, Cear\'a, Brazil}
\affiliation{Computational Physics for Engineering Materials, IfB, ETH
Zurich, Wolfgang-Pauli-Strasse 27, CH-8093 Zurich, Switzerland}
\author{H.~J.~Herrmann}
\email{hans@ifb.baug.ethz.ch}
\affiliation{Computational Physics for Engineering Materials, IfB, ETH Zurich, Wolfgang-Pauli-Strasse 27, CH-8093 Zurich, Switzerland}
\affiliation{Departamento de F\'{\i}sica, Universidade Federal do Cear\'a, 60451-970 Fortaleza, Cear\'a, Brazil}

\begin{abstract}
Two dimensional space-filling bearings are dense packings of disks that can
rotate without slip. We consider the entire first family of bearings for loops
of size four and propose a hierarchical construction of their contact network.
We provide analytic expressions for the clustering coefficient and degree
distribution, revealing bipartite scale-free behavior with tunable degree
exponent depending on the bearing parameters.  We also analyze their average
shortest path and percolation properties.
\end{abstract}

\maketitle
%
\section{Introduction}
Bearings are mechanical dissipative systems of rotors that, when perturbed,
relax towards a bearing state, where all touching rotors rotate without slip.
When these bearings cover the entire space they are called space-filling
bearings~\cite{Herrmann90}. Moreover,if such packings are sheared between
moving surfaces, they can be used as a model to explain the existence of
regions where tectonic plates can creep on each other for long periods of time
without triggering earthquake activity, known as seismic gaps~\cite{Lomnitz82}.
Space-filling bearings have also been used as a heuristic model for scale-free
velocity fields, where the superdiffusion of massive particles can take place
\cite{Baram10}.

Herrmann \textit{et al.}~\cite{Herrmann90} presented a numerical algorithm to
construct configurations of two dimensional space-filling bearings of
polydisperse disks for loops of size four on a stripe geometry.  They showed
that two families of bearings can be obtained, where each configuration is
classified by two integer indices $m$ and $n$. The contact network of a bearing
is obtained by mapping it into a graph, where nodes are the disks and links are
established between touching disks. In the bearing state, which has no slip,
two disks rolling on each other must have opposite sense of rotation. The
contact networks are thus bipartite, with the type of node defined by its sense
of rotation. The topological properties of the contact network are intimately
related to the force chains~\cite{Hidalgo02} and the dynamical response of the
bearing to perturbations~\cite{Araujo13}.

Andrade \textit{et al.}~\cite{Andrade05} have shown that the contact network of
Apollonian packings is a scale-free, small world, Euclidean, space-filling, and
matching graph.  The interesting properties of this network, named Apollonian
network, have motivated a series of follow-ups to study their
geometrical~\cite{Doye05}, magnetic~\cite{Andrade05b,Andrade09,Araujo10},
spectral~\cite{Andrade05c}, and dynamical
properties~\cite{Pellegrini07,Zhang09}. Even an extension to random networks
has been proposed~\cite{Zhang07}. In contrast to bearings where loops are
necessarily of an even number of disks, for the Apollonian network loops are of
size three.

Here, we consider the first family of space-filling bearings in the
stripe geometry and analyze their contact networks. Doye and Massen
looked at these networks in the limit $m=n$ and provided heuristic
arguments to estimate their degree exponent~\cite{Doye05}. We propose a
hierarchical construction of such networks which allows us to analyze
the entire range of indices $m$ and $n$ and provide analytic expressions
for the degree distribution and clustering coefficient. We also describe
several other properties. The paper is organized as follows. In
Section~\ref{sec::model_definition} we start with the special case
$m=n=0$. The general case is discussed in Section~\ref{sec::general}. We
finally draw some conclusions in Section~\ref{sec::final_remarks}.

\section{\label{sec::model_definition}Network for $m=n=0$}
\subsection{\label{sec::definition_special_case}The network construction}
\begin{figure}
\includegraphics[width=\columnwidth]{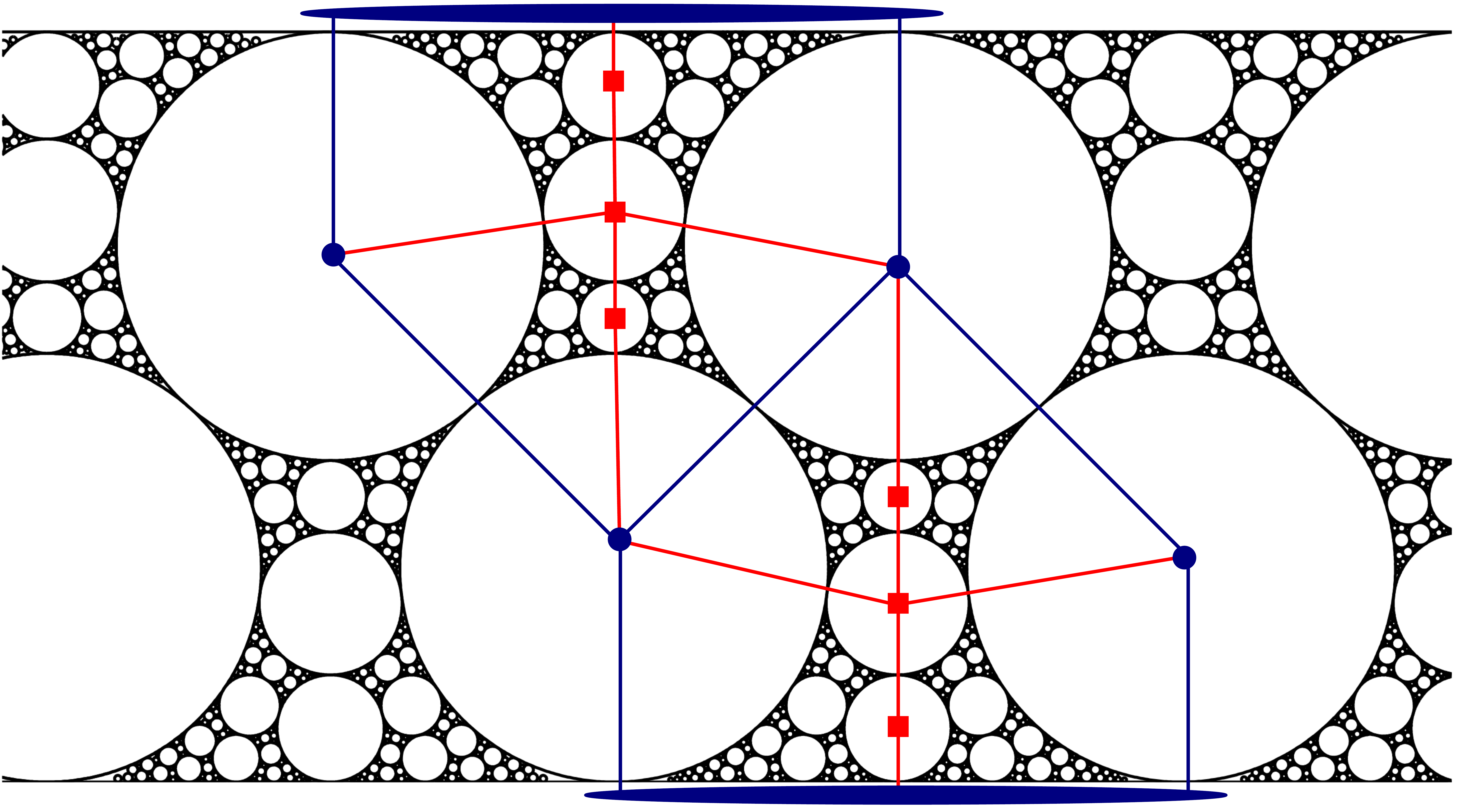}
\caption{\label{fig::stripe_n_0_m_0} (Color online) Section of the $m,n=0$
bearing and the first generation of its contact network.}
\end{figure}

\begin{figure*}
\includegraphics[width=\textwidth]{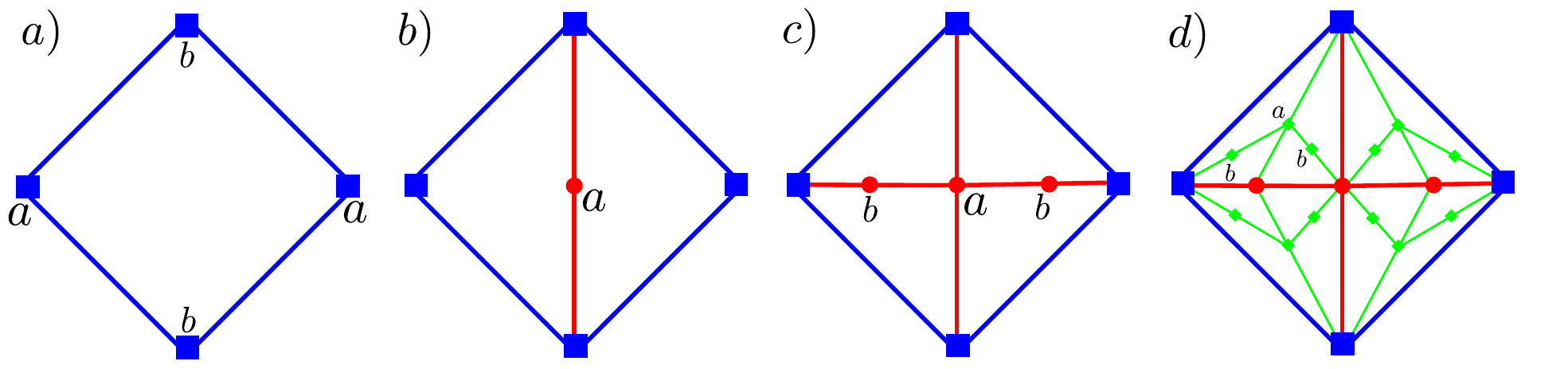}
\caption{\label{fig::gen_n_0_m_0}
Hierarchical rule to construct the contact network for $m=n=0$. a) One starts
with a loop of four nodes (two $a$- and two $b$-nodes) in a loop arrangement.
b) A new $a$-node is added to the center of the loop and connected to the
$b$-nodes on the top and bottom of the loop, dividing it into two loops. c) Two
new $b$-nodes are also added, one inside of each loop and connected to the two
closest $a$-nodes. The initial loop is now divided into four equal loops. d)
The new generation is constructed in the same way inside each loop.
}
\end{figure*}

We begin with the specific case of the space-filling bearing of $m=n=0$ (see
Fig.~\ref{fig::stripe_n_0_m_0}). This bearing has translational symmetry with a
unit-cell composed of two topologically identical loops of four disks, defined
by the largest disks, where the top and bottom surfaces are treated as disks of
infinite radius.  For $m=n$ the bearing has also $C2$ rotation symmetry around
the center of the common edge of the two largest loops. Thus, it is sufficient
to consider the hierarchical construction rule for the contact network of one
loop in the unit cell. By construction, all loops consist of an even number of
disks and the network is bipartite, with two types of nodes denoted as $a$- and
$b$-nodes. The construction rule is summarized in Fig.~\ref{fig::gen_n_0_m_0}.
One starts with a loop arrangement of four nodes (two $a$- and two $b$-nodes),
corresponding to the four disks of the largest loop. An $a$-node is only
connected to $b$-nodes. The first generation $g=1$ is constructed by adding an
$a$-node to the center of the loop and connecting it to the two $b$-nodes,
splitting the loop into two loops. This new $a$-node corresponds to the central
disk touching the two lateral ones in each loop of the unit cell shown in
Fig.~\ref{fig::stripe_n_0_m_0}.  Inside each loop one $b$-node is included and
connected to the two closest $a$-nodes. These two new nodes correspond to the
other two disks vertically aligned with the previous one. At the end, the
initial square is divided into four loops. The next generations are obtained
hierarchically by repeating the same procedure inside each loop. By
construction, the contact network is planar and self-similar.

\subsection{\label{sec::dec_dist} Degree distribution} 
We now provide an analytic expression for the degree distribution $P(k)$, where
$k$ is the node degree (number of touching disks). Let us start with the number
of nodes $N(g)$ at generation $g$, and neglect the first four nodes. One starts
with one loop of two $a$- and two $b$-nodes at generation
zero. At each generation, each loop is divided into four. Thus, the
final number of loops is $4^g$.  For each loop in generation
$g-1$, one $a$- and two $b$-nodes are added to obtain the generation $g$, so
that the number of $a$-nodes $N_a$ changes from generation $g-1$ to $g$ as,
\begin{equation} \label{eq::delta_node_a}
\Delta N_a(g)=4^{g-1}
\end{equation}
and the number of $b$-nodes $N_b$ as,
\begin{equation} \label{eq::delta_node_b}
\Delta N_b(g)=2\cdot4^{g-1} \ \ .
\end{equation}
The number of nodes at generation $g$ is then,
\begin{equation}
N_a(g)=\sum_{t=1}^{g} 4^{t-1}=\frac{4^g-1}{3}
\end{equation}
and
\begin{equation}
N_b(g)=\sum_{t=1}^{g} 2\cdot 4^{t-1}=\frac{2(4^g-1)}{3} \ \ ,
\end{equation}
respectively. The total number of nodes $N$ is
\begin{equation}
N(g)=N_a(g)+N_b(g)=4^g-1 \ \ .
\end{equation}

At each generation, all $a$- and $b$-nodes receive one new link for
each adjacent loop. Since the number of such loops
equals the degree, the latter doubles at each generation. The new
$a$-nodes have degree four, while the $b$-nodes have degree two. Hence,
at generation $g$, the degree $k(t)$ of a node, added at generation
$g_0$ that is part of the network for $t=g-g_0$ generations is,
\begin{equation}\label{eq::degree_gen_a}
k_a(g-g_0)=4\cdot 2^{g-g_0} \ \ ,
\end{equation}
\begin{equation}\label{eq::degree_gen_b}
k_b(g-g_0)=2\cdot 2^{g-g_0} \ \ .
\end{equation}

At generation $g$ the degree of a node is related to the generation $g_0$ at
which the node was added. This generation is given by,
\begin{equation}
g-g_{0a}(k)=\frac{\ln(k/2)}{\ln 2}-1 \ \ , 
\end{equation}
and
\begin{equation}
g-g_{0b}(k)=\frac{\ln(k/2)}{\ln 2} \ \ .
\end{equation}
The number of nodes of degree $k$ at generation $g$ equals the number of nodes
added at generation $g_0(k)$, given by
Eqs.~(\ref{eq::delta_node_a})~and~(\ref{eq::delta_node_b}). Thus, the degree
distribution $P_a(k,g)$ is,
\begin{eqnarray}
P_a(k,g) &=& \frac{\Delta N_a(g_{0a}(k))}{N_a(g)} \nonumber \\
 &=& 3\frac{4^g}{4^g-1}\left(\frac{k}{2}\right)^{-2} .
\end{eqnarray}
In the same way, $P_b(k,g)$ is,
\begin{eqnarray}
P_b(k,g) &=& \frac{\Delta N_b(g_{0b}(k))}{N_b(g)} \nonumber \\
 &=& \frac{3}{4}\frac{4^g}{4^g-1}\left(\frac{k}{2}\right)^{-2} .
\end{eqnarray}
The total degree distribution $P(k,g)$ is then,
\begin{eqnarray}
P(k,g) &=& \frac{\Delta N_a(g_{0a}(k))+\Delta N_b(g_{0b}(k))}{N(g)} \nonumber
\\
 &=& \frac{3}{2}\frac{4^g}{4^g-1}\left(\frac{k}{2}\right)^{-2} .
\end{eqnarray}
In the limit $g\rightarrow \infty$,
\begin{eqnarray}
P_a(k) &=& 3\left(\frac{k}{2}\right)^{-2} \ \ , \\
P_b(k) &=& \frac{3}{4}\left(\frac{k}{2}\right)^{-2} \ \ , \\
P(k) &=& \frac{3}{2}\left(\frac{k}{2}\right)^{-2} \ \ .
\end{eqnarray}
Thus, the degree distribution scales as $P(k)\propto k^{-\gamma}$, with
$\gamma=2$, corresponding to a scale-free network. This exponent is larger
than the one obtained for the Apollonian network, where $\gamma=\frac{\ln
3}{\ln 2}\approx 1.585$~\cite{Andrade05}. Note that the $a$/$b$ asymmetry
disappears when one considers the two topological identical loops in the entire
unit cell, for the $a$-nodes in the top loop correspond to the $b$-nodes in the
bottom one.

\subsection{\label{sec::small_world}Shortest paths and clustering coefficient}
\begin{figure}
\includegraphics[width=\columnwidth]{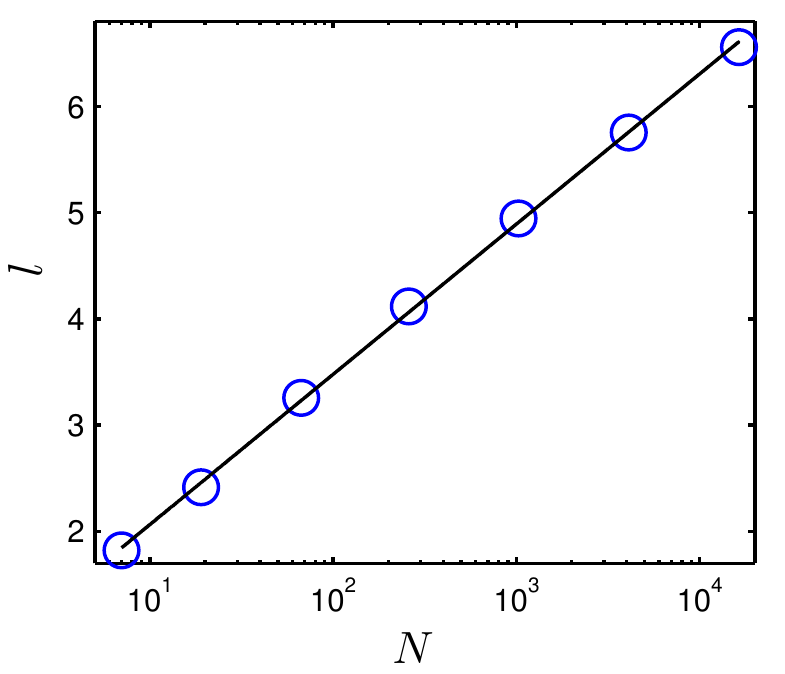}
\caption{\label{fig::l_n_0_m_0}Size dependence of the average shortest path
$l$, for the contact network of bearings of $m=n=0$, where $N$ is the number of
nodes. The shortest path scales logarithmically with the number of nodes,
$l=0.61\ln{N}+0.65$ as expected for small-world networks.}
\end{figure}

Spatial, self-similar networks are expected to exhibit some form of small-world
behavior due to the confinement of connections~\cite{Doye05}, which is, for
example, the case of the Apollonian network~\cite{Zhang08}.  Numerically, this
can be checked by analyzing the size dependence of the average shortest path
$l$, defined as the average minimum number of links necessary to form a
connecting path between pairs of nodes in the network.
Figure~\ref{fig::l_n_0_m_0} shows that $l=a\ln N+b$, where $a$ and $b$ are
constants, corresponding to a logarithmic scaling, as expected for small-world
networks~\cite{Watts98}. 

\begin{figure}
\includegraphics[width=\columnwidth]{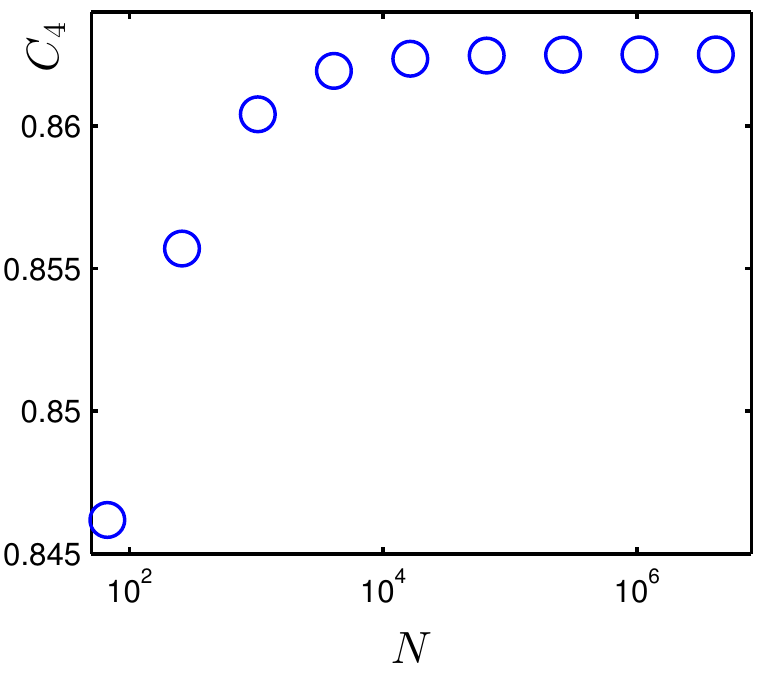}
\caption{\label{fig::C_n_0_m_0}Clustering coefficient $C_4$ for the contact
network of bearings of $m=n=0$ as a function of the number of nodes $N$.}
\end{figure}

Small-world networks are typically highly clustered~\cite{Watts98}. To quantify
the degree of clustering one measures the clustering coefficient $C$, defined
for each node as the fraction of pairs of neighbors that are directly
connected, forming a triangle. In the case of bearings, all loops have a even
number of nodes, the contact network is bipartite, and two neighbors of a node
are never directly connected.  Lind~\textit{et al.}~\cite{Lind05} proposed a
new clustering coefficient for bipartite networks $C_4$, defined as the
fraction of pairs that are indirectly connected through one single node,
forming a loop.  Then, for a node of degree $k$,
\begin{equation}
C_4=\frac{\#(\text{Indir. conn. between neighb.})}{k(k-1)/2}.
\end{equation}
In the following we will use this definition. First, we calculate
$C_{4a/b}(t)$, the clustering coefficient of an $a$/$b$-node that was added to
the network $t$ generations before. At each iteration, the degree of every node
is doubled, by adding new neighbors. Each new neighbor is connected via a new
node to other two neighbors (see Fig.~\ref{fig::gen_n_0_m_0}). Thus, the number
of indirectly connected pairs of neighbors increases at each generation by
twice the node degree. Every new $a$-node has degree four and from its six
different pairs of neighbors, five are indirectly connected. Every new $b$-node
has degree two and its pair of neighbors is always indirectly connected.  By
summing over generations, one gets for $a$-nodes,
\begin{eqnarray}
C_{4a}(t)&=&\frac{5+\sum_{i=0}^{t-1}2\cdot k_a(i)}{k_a(t)(k_a(t)-1)/2}
\nonumber \\ &=&\frac{3}{2}2^{-t}-\frac{2}{4\cdot 2^t-1},
\end{eqnarray}
and for $b$-nodes,
\begin{eqnarray}
C_{4b}(t)&=&\frac{1+\sum_{i=0}^{t-1}2\cdot k_b(i)}{k_b(t)(k_b(t)-1)/2}
\nonumber \\ &=&3\cdot 2^{-t}-\frac{2}{2\cdot 2^t-1},
\end{eqnarray} 
where we employed Eqs.~(\ref{eq::degree_gen_a})~and~(\ref{eq::degree_gen_b}).
Once again, note that the $a$/$b$ asymmetry is only observed when we solely
consider one loop.  For the entire stripe, the top and bottom loops of the unit
cell are equivalent, but the $a$-nodes of the top loop are $b$-nodes of the
bottom one. Thus, when the entire unit cell is considered, the network is
completely symmetric with respect to $a$ and $b$.  Both coefficients tend to
zero as $C_4(t)\sim 2^{-t}$ for $t\rightarrow \infty$.  The clustering of the
entire network can be evaluated from the average over all nodes,
\begin{equation}
C_4(g)=\sum_{t=1}^{g}\frac{\Delta N_a(t)C_{4a}(g-t)+\Delta
N_b(t)C_{4b}(g-t)}{N(g)}.
\end{equation}
We evaluate this sum numerically as shown in Fig.~\ref{fig::C_n_0_m_0}, for
different number of nodes in the network, and obtain that $C_4\approx0.8625$ in
the thermodynamic limit.

\subsection{\label{sec::spec_percol}Bond percolation}
\begin{figure}
\includegraphics[width=\columnwidth]{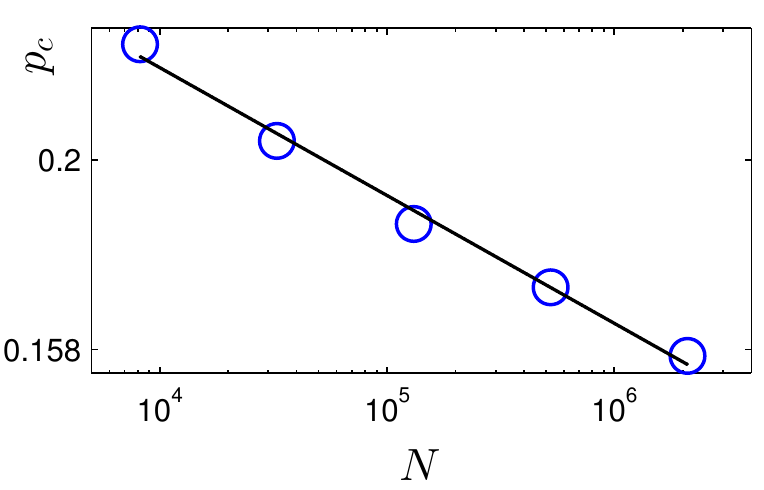}
\caption{\label{fig::pc_n_0_m_0} Dependence o the percolation threshold $p_c$
on the network size $N$. The black line corresponds to the least-squares fit to
the data of a power-law, $p_c\sim N^{-frac{1}{2\nu}}$, with exponent
$\nu=7\pm1$.}
\end{figure}

We now study bond percolation on a network corresponding to a unit-cell of the
stripe, consisting of two initial squares sharing one link (see
Fig.~\ref{fig::stripe_n_0_m_0}). We focus on the existence of a spanning
cluster between the two nodes representing the top and bottom surfaces,
respectively. To compute the percolation threshold, we performed Monte Carlo
simulations for different values of bond occupation probability $p$ and network
size $N$. We estimate the threshold $p_c$ as the value of $p$ at which the
probability of spanning is $1/2$. We employed the bisection method and
considered values of $p$ that differ by $0.001$. Figure~\ref{fig::pc_n_0_m_0}
shows the size dependence of the estimated value of $p_c$, where one clearly
sees that $p_c$ vanishes in the thermodynamic limit. Asymptotically, the decay
follows a power-law $p_c\sim N^{-\frac{1}{2\nu}}$, with $\nu\approx7$. The
same threshold is observed for the Apollonian network and other scale-free
networks with $\gamma<3$. However, the convergence to the thermodynamic value
is much slower here than for the Apollonian network, where
$\nu\approx3$~\cite{Andrade05}.

\section{\label{sec::general}General case}

\subsection{\label{sec::net_gen}The network construction}
\begin{figure*}
  \begin{tabular}{c | c | c}
    $m=0,n=0$ & $m=0,n=1$ & $m=0,n=2$\\
    \includegraphics[width=0.37\textwidth]{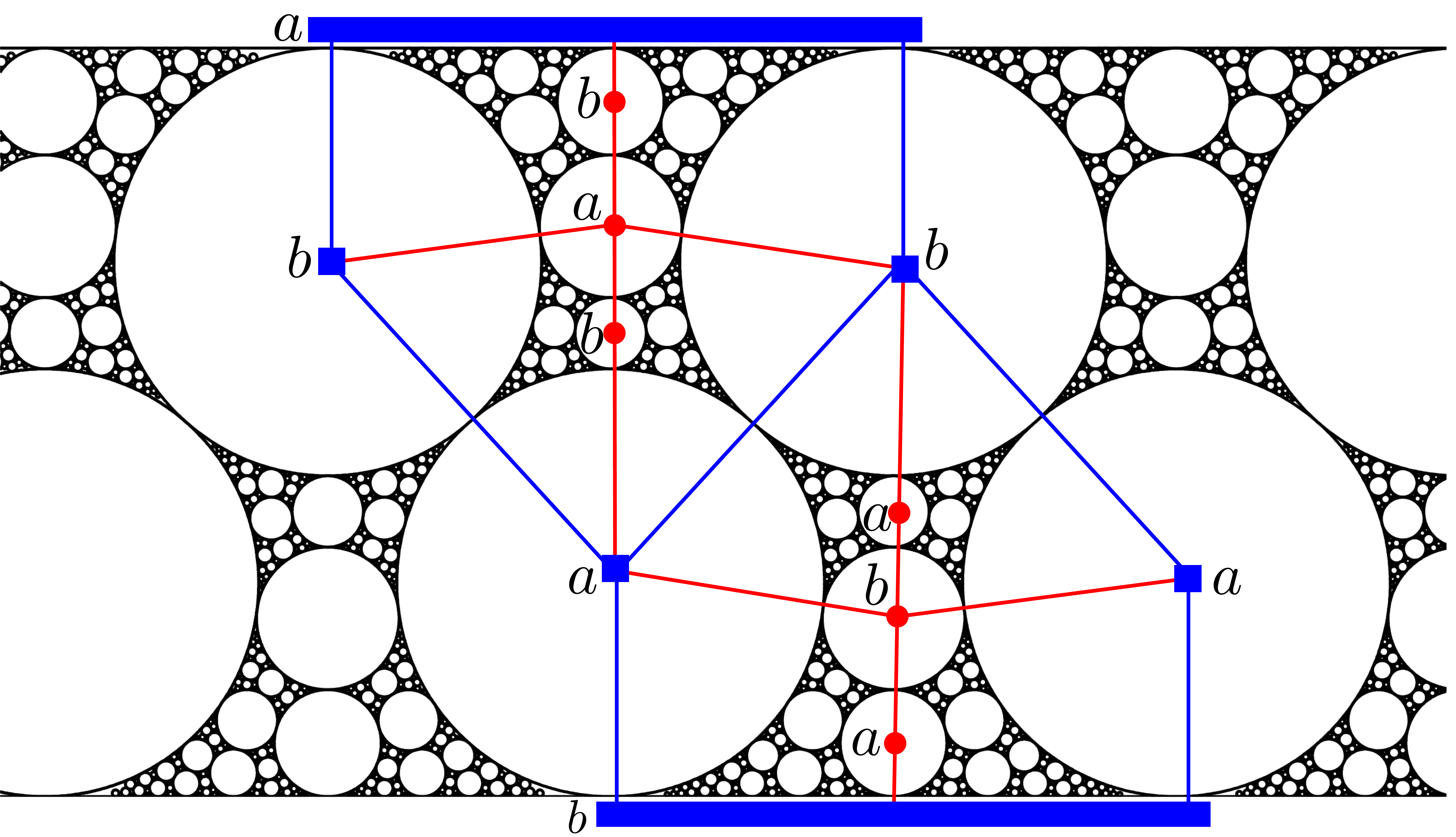} &
    \includegraphics[width=0.305\textwidth]{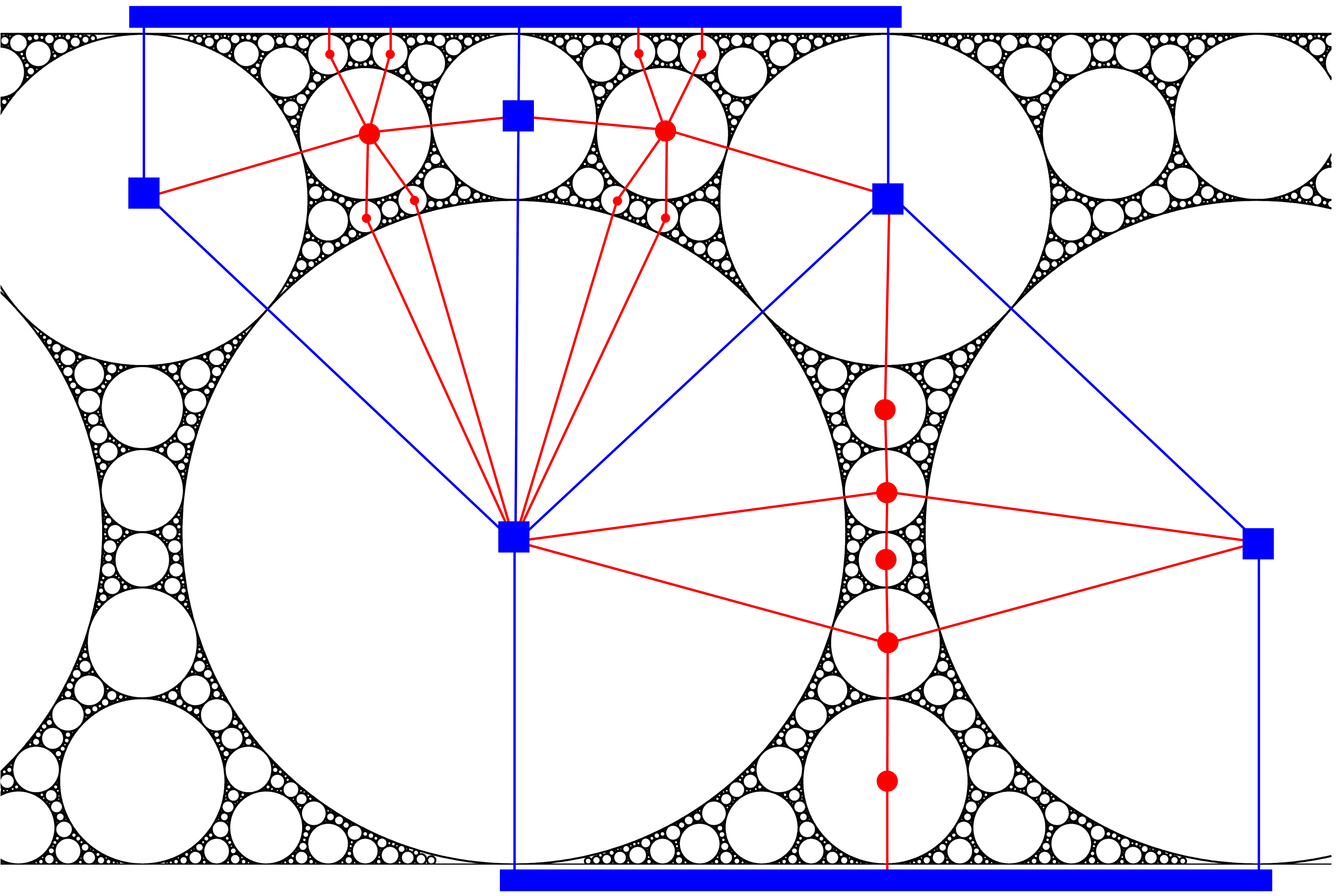} &
    \includegraphics[width=0.30\textwidth]{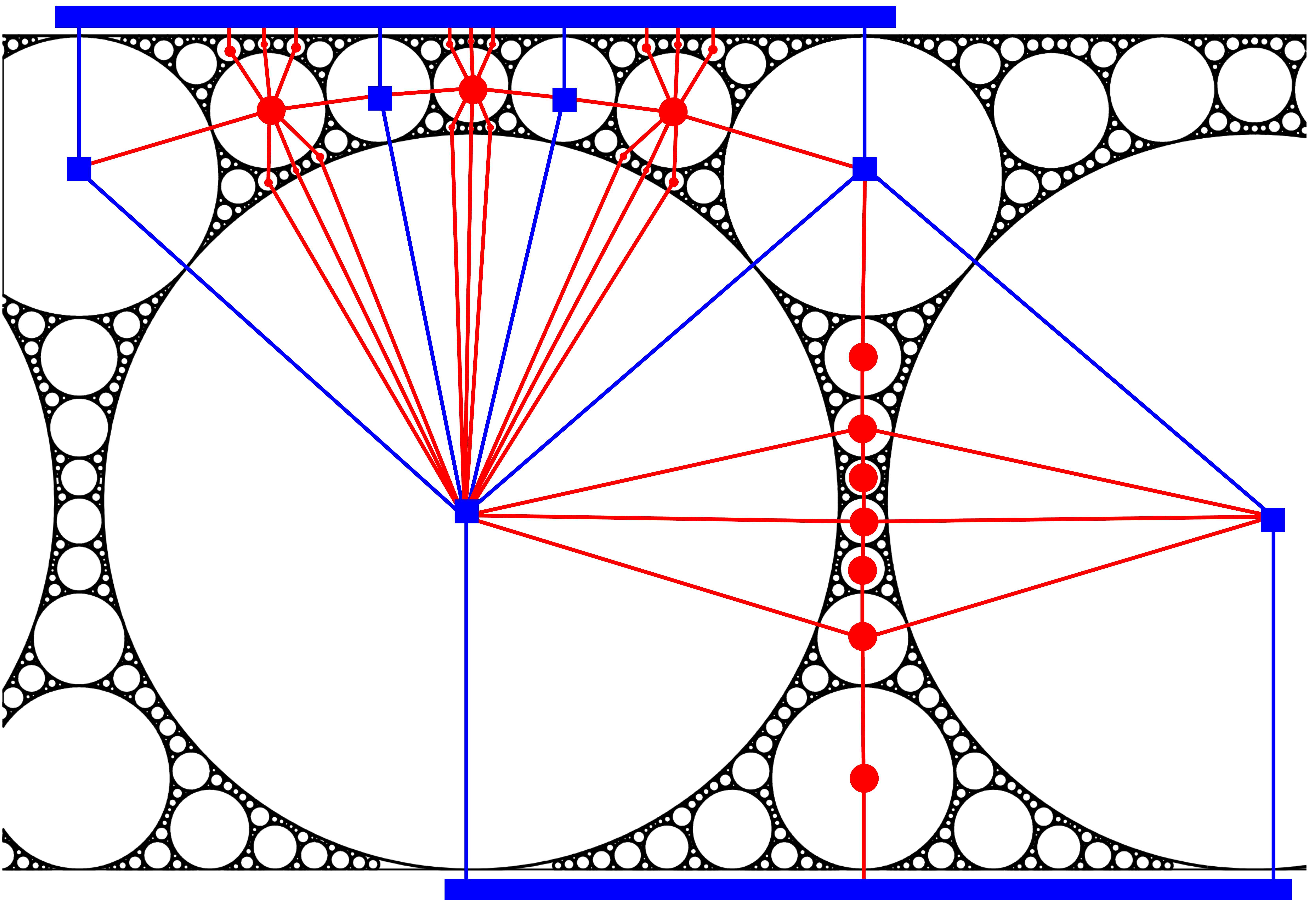}\\
    $m=1,n=0$ & $m=1,n=1$ & $m=1,n=2$\\
    \includegraphics[width=0.34\textwidth]{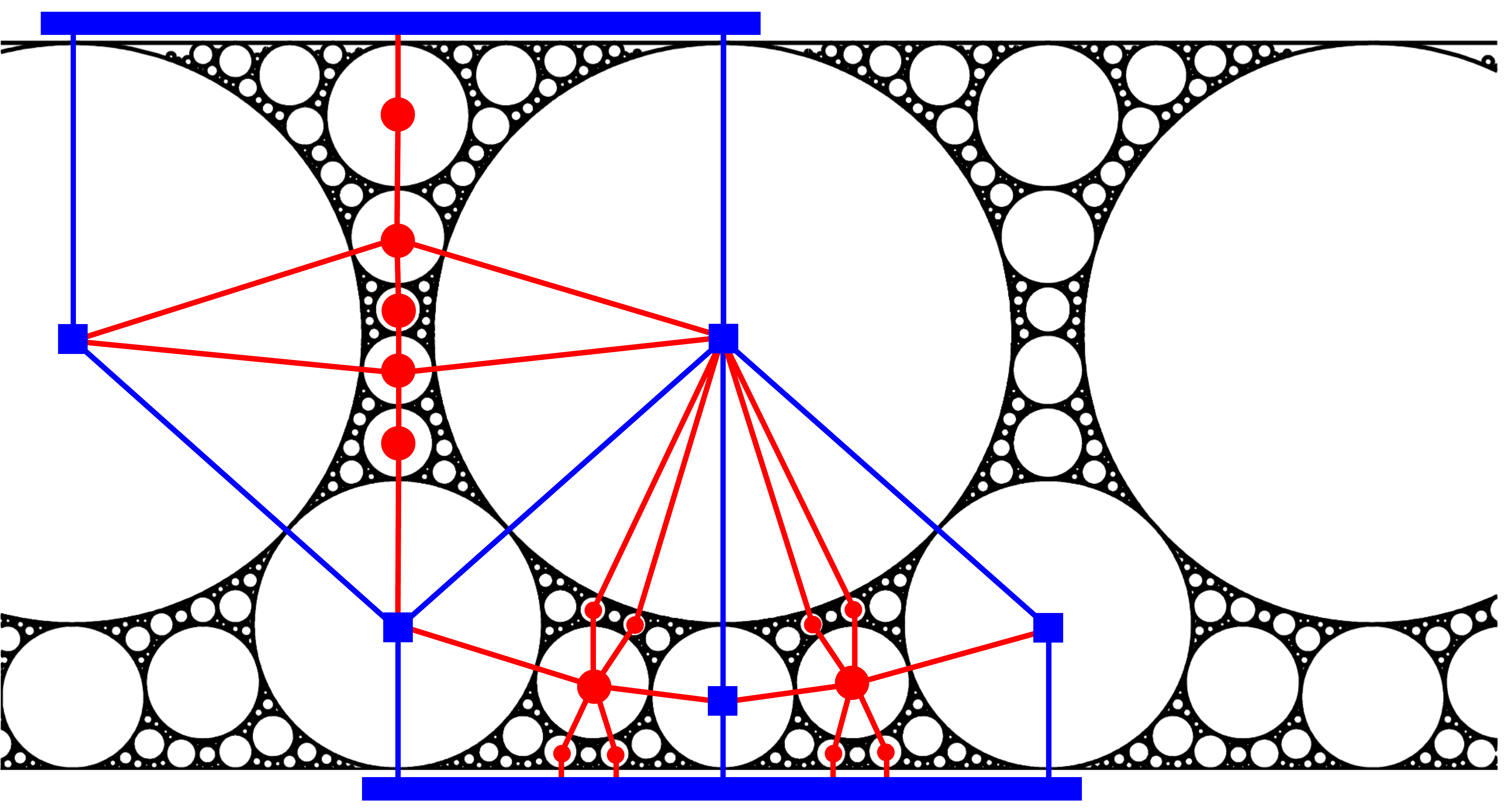} &
    \includegraphics[width=0.31\textwidth]{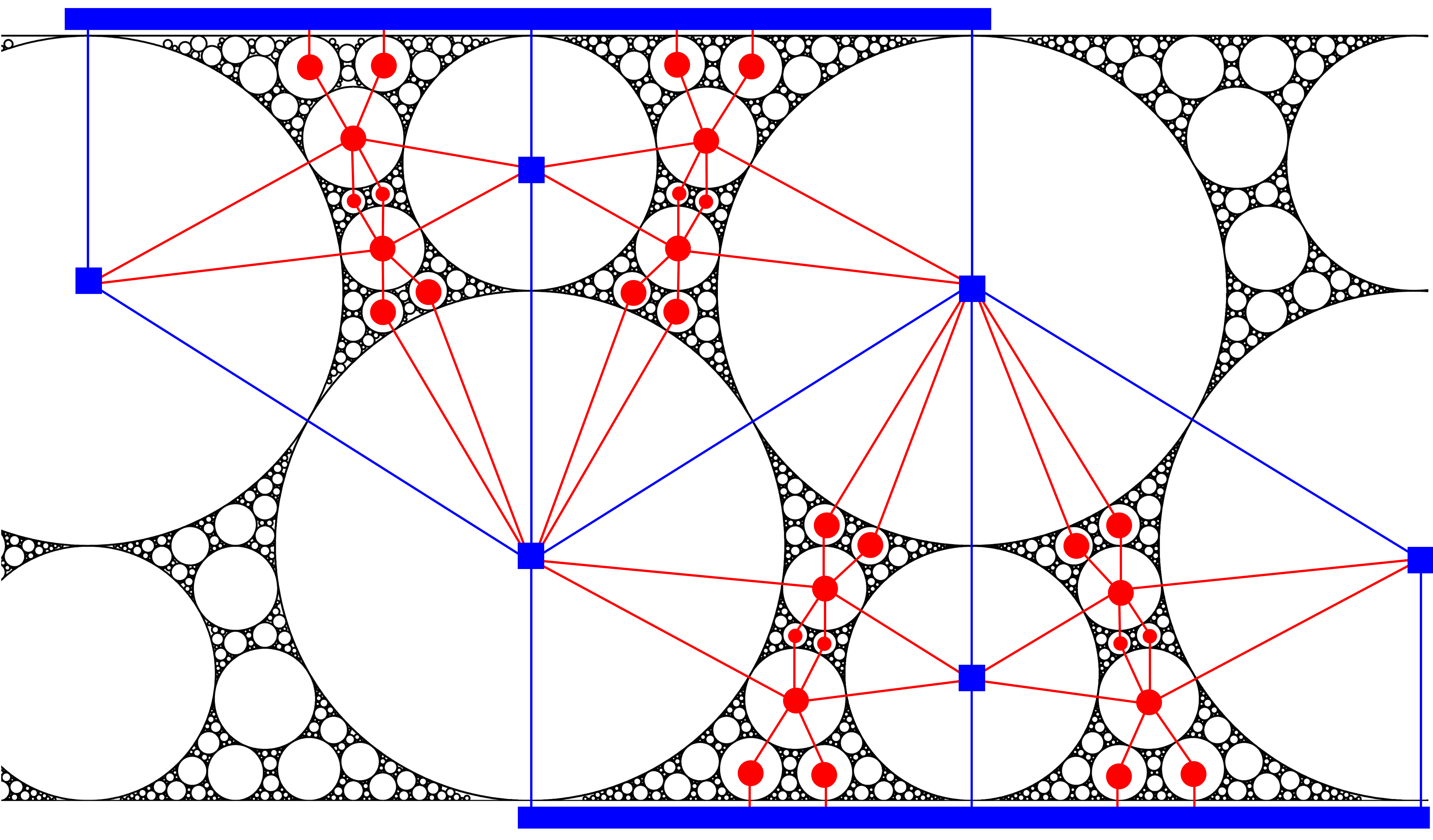} &
    \includegraphics[width=0.34\textwidth]{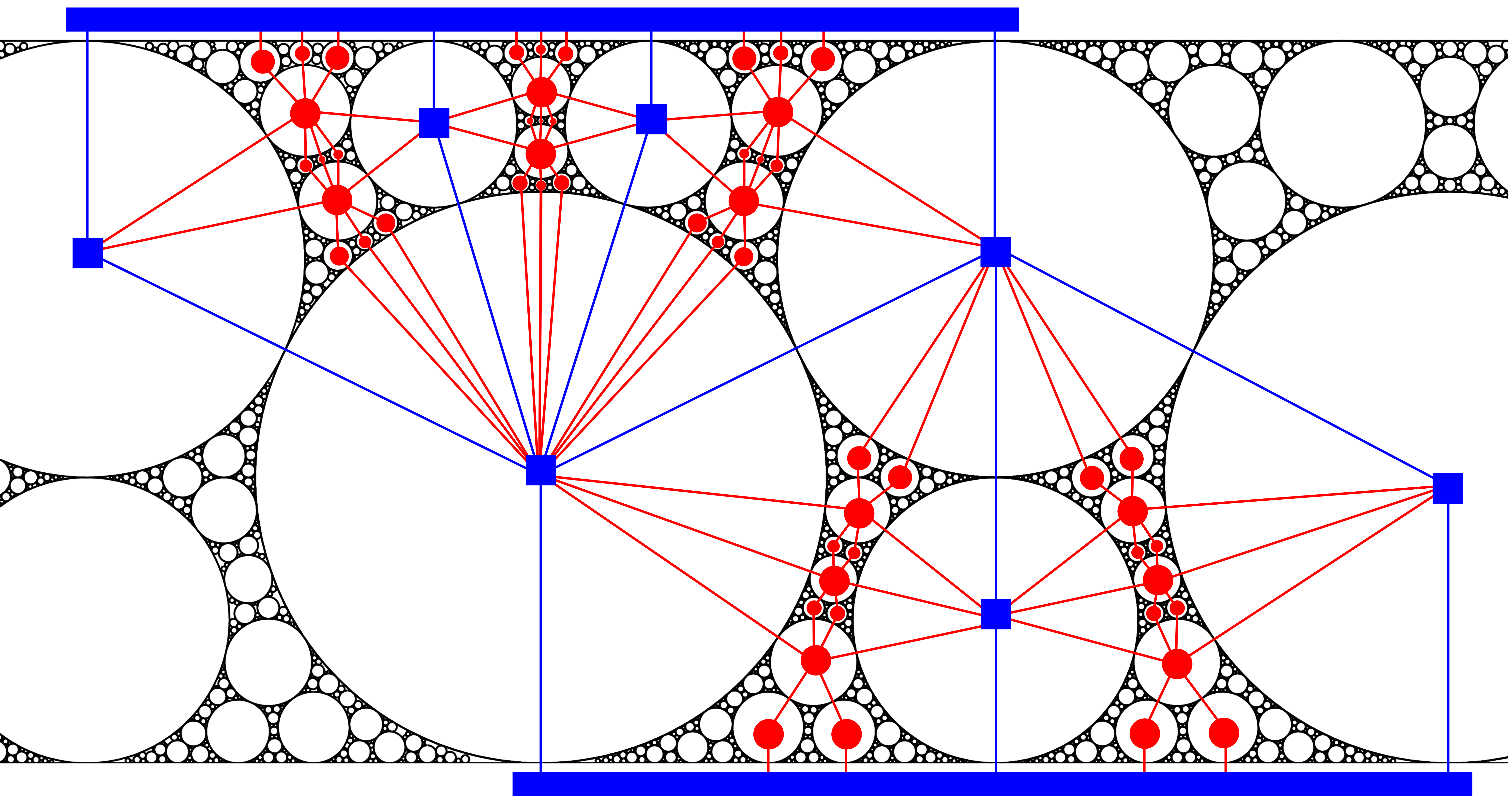}\\
    $m=3,n=0$ & $m=3,n=1$ & $m=3,n=2$\\
    \includegraphics[width=0.32\textwidth]{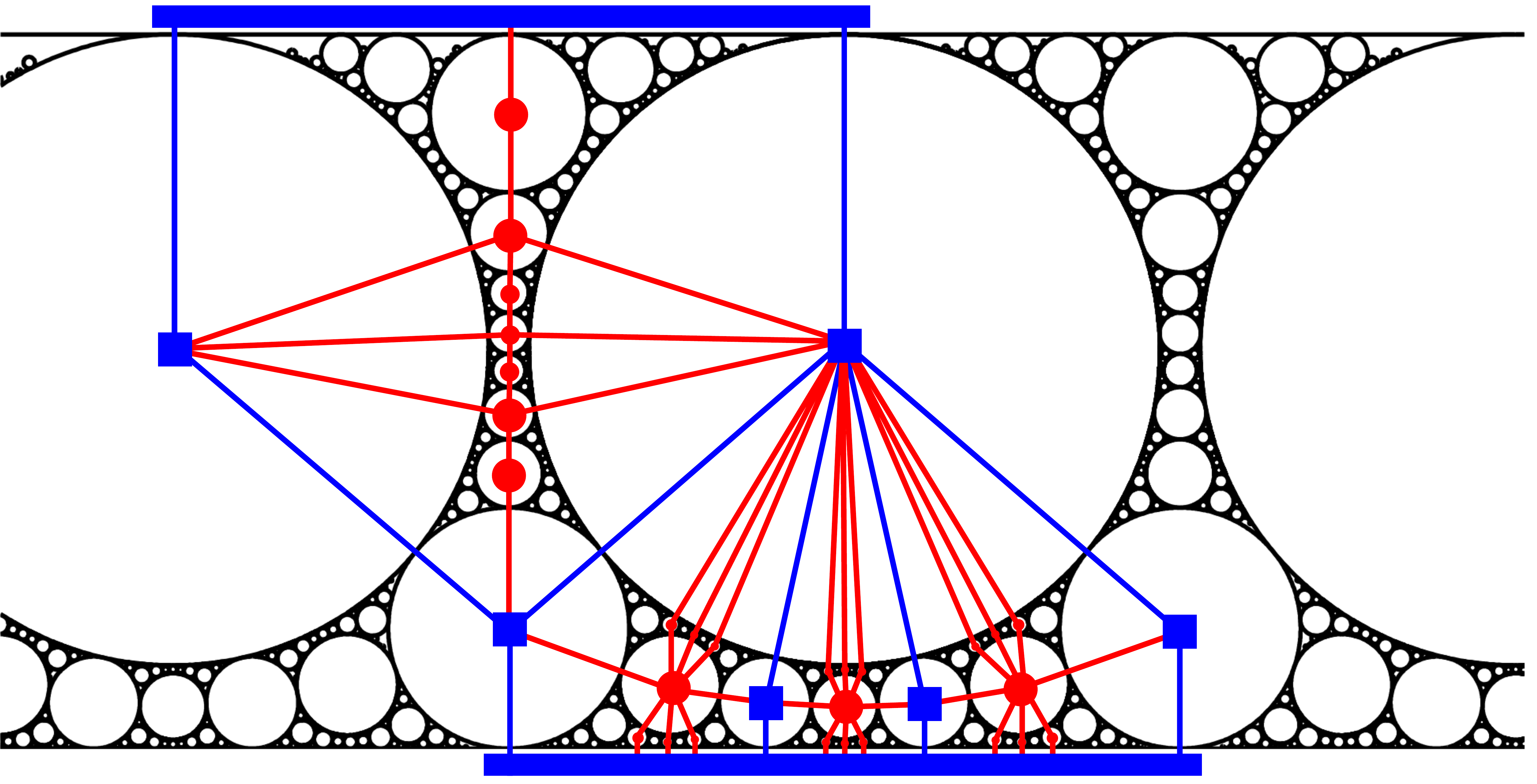} &
    \includegraphics[width=0.293\textwidth]{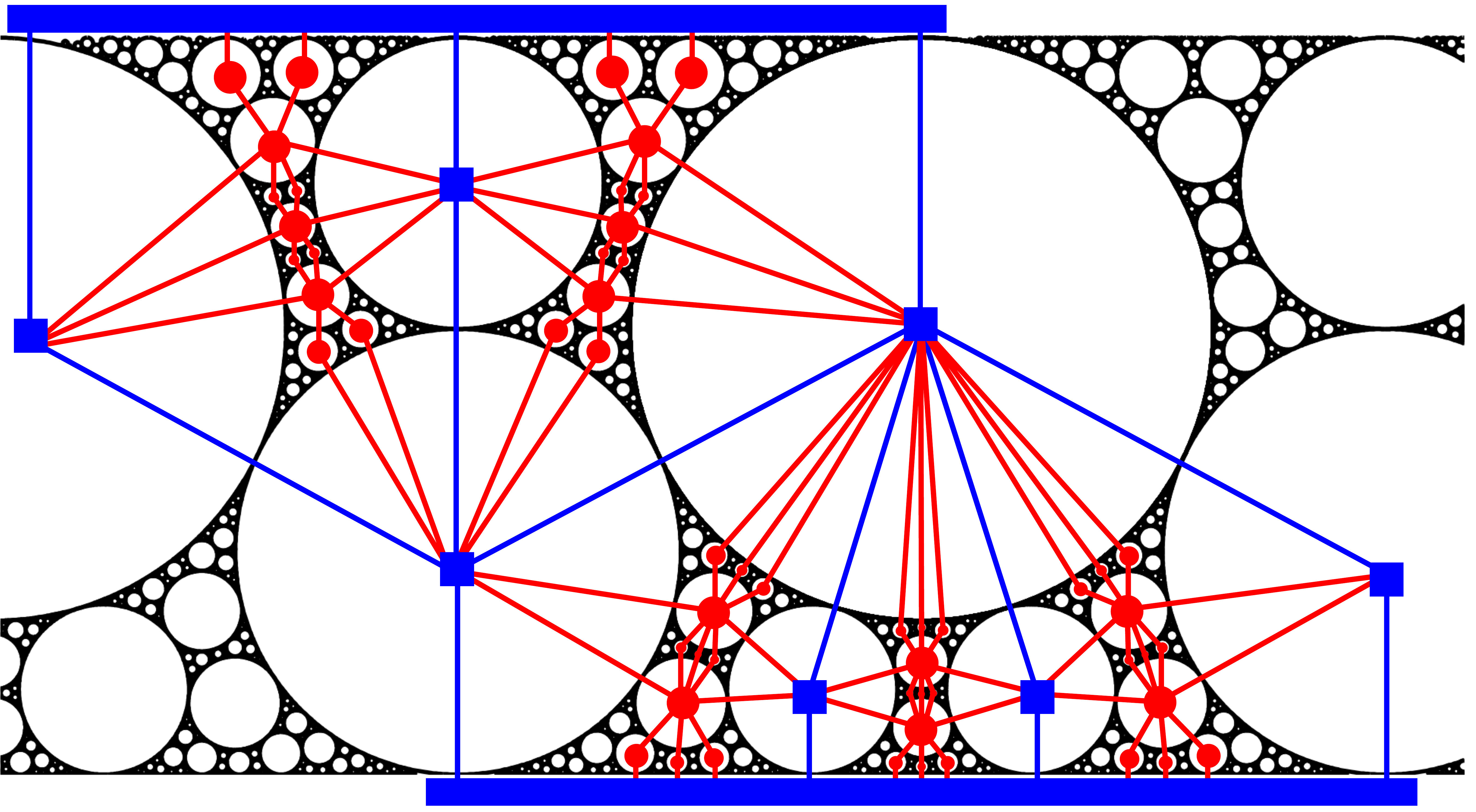} &
    \includegraphics[width=0.35\textwidth]{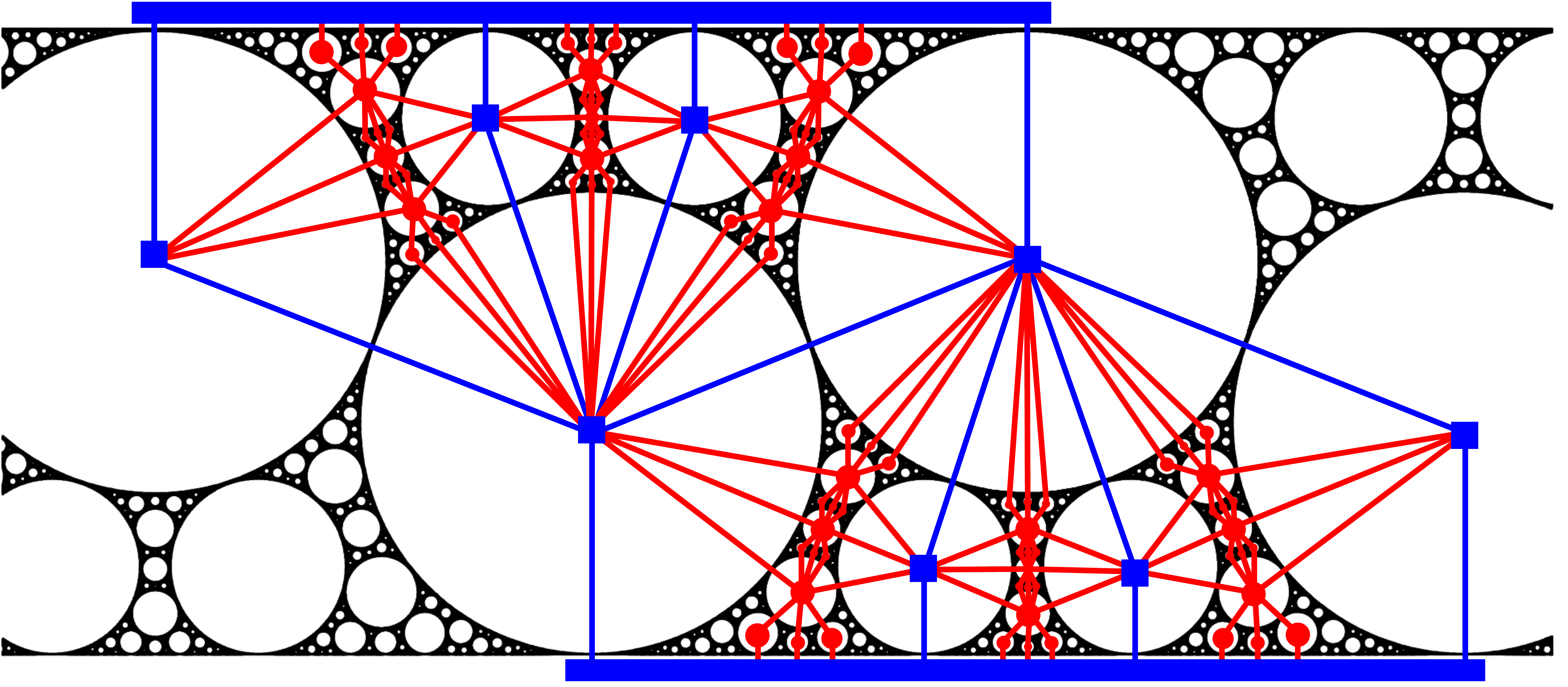}
  \end{tabular}
  \caption{\label{fig::gaskets_stripe}(Color online) Examples of space-filling
bearings of the first family, for different values of $m$ and $n$. }
\end{figure*}
\begin{figure*}
\includegraphics[width=\textwidth]{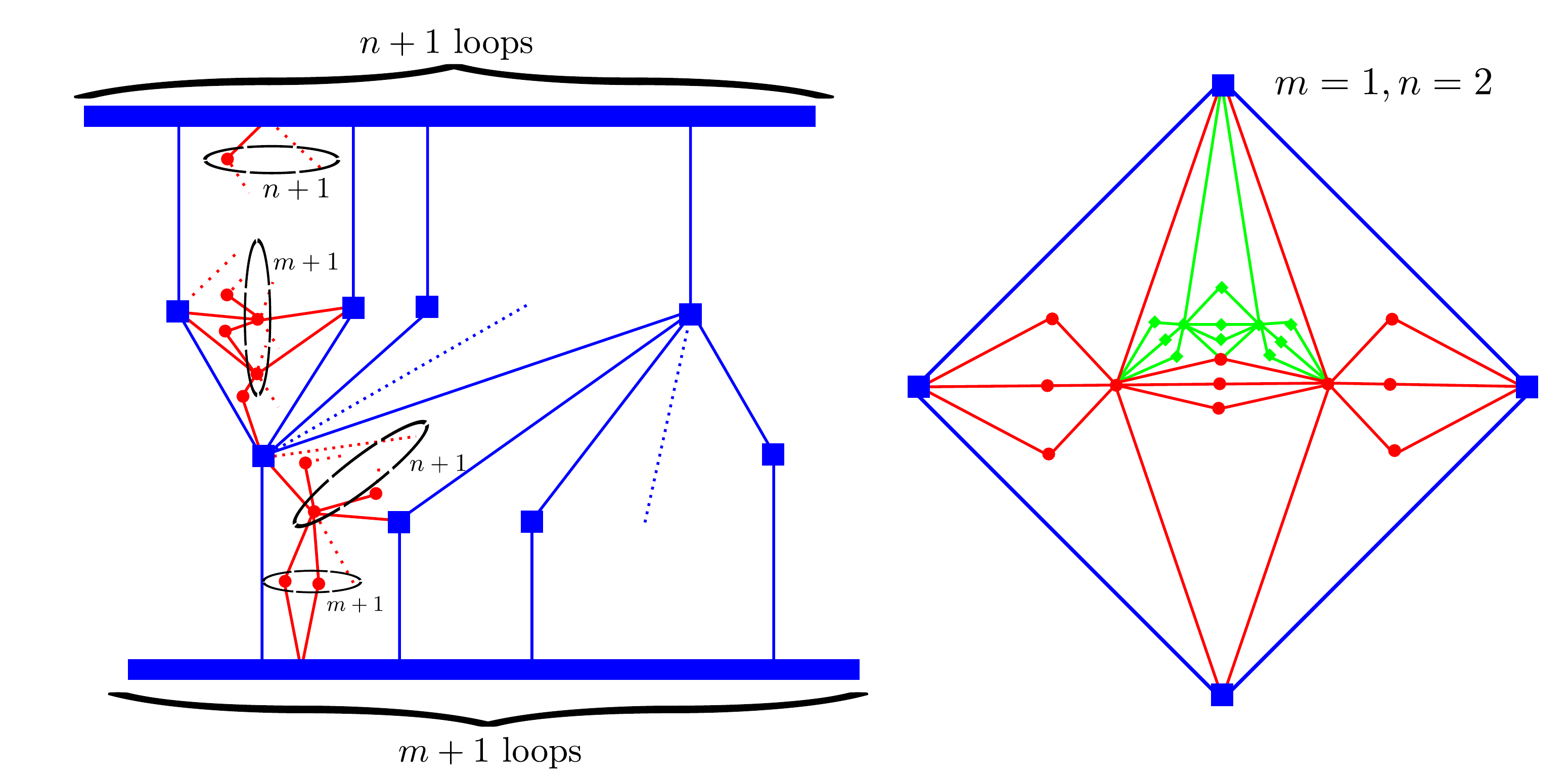}
\caption{\label{fig::net_gen_gen}(Color online) The left panel shows a sketch
of the initial network set up (blue, sites drawn as squares) and the first
generation in the general case (red, sites as dots). In the right panel are the
first two generations and part of the second (green) for $m=1$ and $n=2$.}
\end{figure*}

We now consider the general case of the contact network for a bearing in the
first family for loops of size four with any $m$ and $n$.
Figure~\ref{fig::gaskets_stripe} shows examples of bearings generated with
different $m$ and $n$, with the respective contact network on top. For all
cases, the unit cell consists of loops of size four with the largest disks,
including the top and bottom surfaces, respectively. However, the number of
such loops varies with $m$ and $n$ and the rotation symmetry is broken for
$m\not=n$. At each iteration the number of vertical and horizontal new loops
constructed inside each loop also depends on $m$ and $n$, respectively. As
summarized in Fig.~\ref{fig::net_gen_gen}, we first discuss how to determine
the initial number of loops in the unit cell (left panel) and proceed
discussing how to hierarchically fill each loop (right panel).

For all values of $m$ and $n$, the unit cell of the bearing consists of a top
and a bottom part. The number of initial loops on top and bottom equals $n+1$ and
$m+1$, respectively (see loops of blue-square nodes in
Fig.~\ref{fig::gaskets_stripe} and left panel in Fig.~\ref{fig::net_gen_gen}).
Note that the $(m,n)$ configuration is equivalent to the $(n,m)$ configuration
after a rotation of $\pi$ around the point where the common edge of the top and
bottom loops crosses the middle of the stripe. To hierarchically construct the
network one starts with the $n+1$ top and $m+1$ bottom loops. Hereafter, we
solely consider one top and a bottom loop (sharing one edge), as the
construction of the other loops is straightforward. To form the first
generation we first add $m+1$ $a$-nodes to the top loop and connect them to the
two existing (lateral) $b$-nodes, dividing the initial loop into $m+2$ loops.
Then, in each new loop, we add $n+1$ $b$-nodes and connect them to the top and
bottom $a$-nodes. We are left with $(n+2)(m+2)$ loops inside the top loop.
Second, we construct the interior of the bottom loop. There, we start by adding
$n+1$ $b$-nodes and connect them to the two existing (lateral) $a$-nodes. Then,
in each one of the new $n+2$ loops we add $m+1$ $a$-nodes and connect them to
the top and bottom $b$-nodes. In the bottom, we are also left with $(n+2)(m+2)$
loops.  Proceeding iteratively in the same way, we hierarchically construct the
entire contact network of the space-filling bearing, for any $m$ and $n$.

\subsection{\label{sec::dec_dist_gen} Degree distribution}
We now provide an analytic expression for the degree distribution
$P_{m,n}(k)$ for any $m$ and $n$, following the same strategy as for
$m=n=0$ in Sec.~\ref{sec::dec_dist}. For simplicity, we restrict the
calculation to networks inside a single loop (one initial top loop),
as the total degree distribution of the unit cell is straightforwardly obtained
as a weighted average of $P_{m,n}(k)$ and $P_{n,m}(k)$,
corresponding to the degree distribution in the top and bottom loops, where the
statistical weights are given by the initial fraction of top $(n+1)$ and bottom
$(m+1)$ loops, respectively.

At each generation, $(n+2)(m+2)$ loops are constructed for every
loop in the previous generation. As before, we neglect the initial set
of nodes. At generation $g-1$ there are $(n+2)^{g-1}(m+2)^{g-1}$ loops. Thus, 
from generation $g-1$ to $g$, the change in the number of $a$-nodes is,
\begin{equation}
\Delta N_a(g)=(m+1)\left[(n+2)(m+2)\right]^{g-1} \ \ ,
\end{equation}
and in the number of $b$-nodes is,
\begin{equation}
\Delta N_b(g)=(m+2)(n+1)\left[(n+2)(m+2)\right]^{g-1} \ \ ,
\end{equation}
corresponding to $(m+1)$ new $a$- and $(m+2)(n+1)$ new $b$-nodes per
loop. The number of nodes at generation $g$ is then,
\begin{eqnarray}
N_a(g)&=&(m+1)\sum_{t=1}^{g}\left[(n+2)(m+2)\right]^{t-1} \nonumber \\
&=&\frac{(m+1)\left(\left[(n+2)(m+2)\right]^g-1\right)}{(n+2)(m+2)-1},
\end{eqnarray}
and
\begin{eqnarray}
N_b(g)&=&(m+2)(n+1)\sum_{t=1}^{g}\left[(n+2)(m+2)\right]^{t-1} \nonumber \\
&=&\frac{(m+2)(n+1)(\left[(n+2)(m+2)\right]^g-1)}{(n+2)(m+2)-1},
\end{eqnarray}
respectively. The total number of nodes $N$ is,
\begin{equation}
N(g)=N_a(g)+N_b(g)=\left[(n+2)(m+2)\right]^g-1.
\end{equation}

\begin{figure*}
\includegraphics[width=0.53\textwidth]{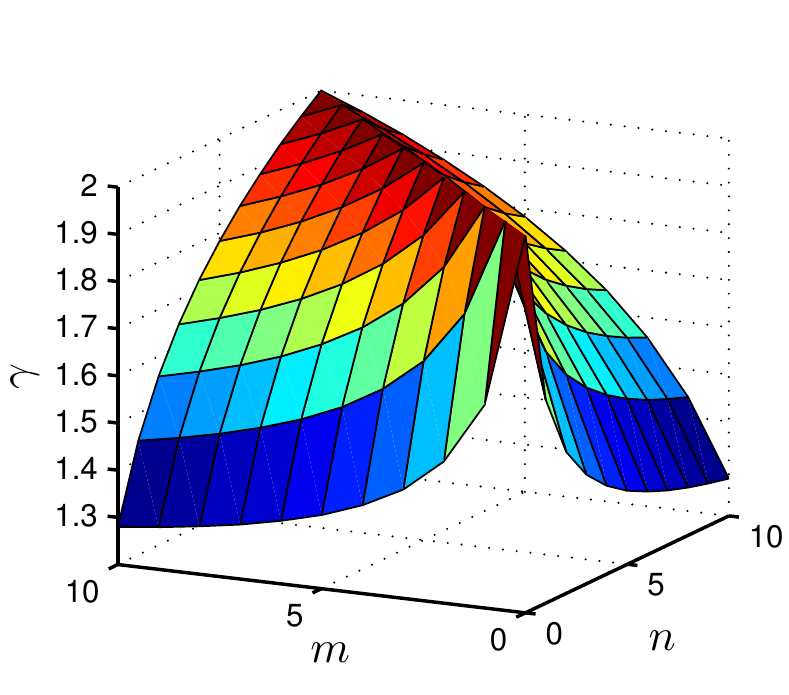}
\includegraphics[width=0.45\textwidth]{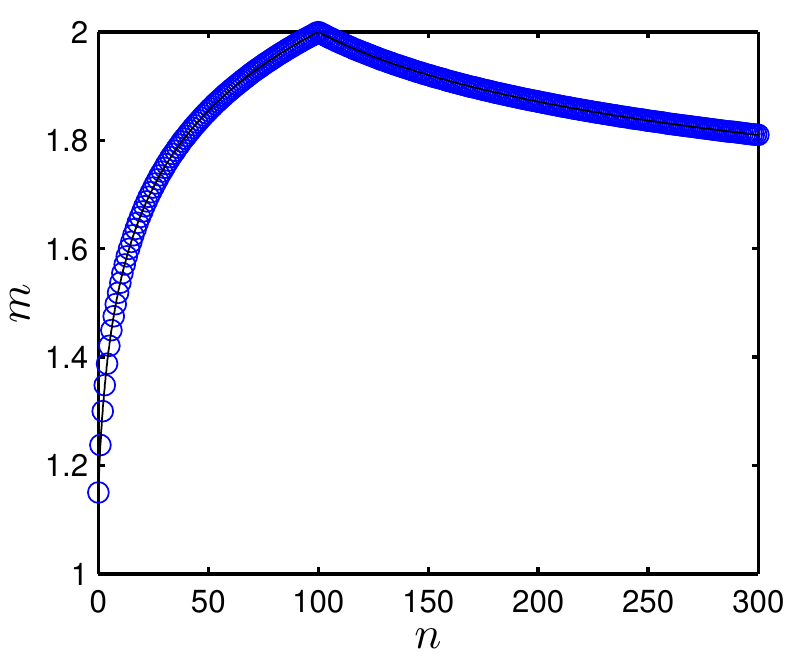}
\caption{\label{fig::gamma} (Color online) Left: Degree exponent
$\gamma$ as a function of $m$ and $n$. Note that, $\gamma=2$ for $m=n$.  Right:
Degree exponent $\gamma$ as a function of $n$ for $m=100$.}
\end{figure*}

The degree $k$ of a node increases monotonically with the generation. An
$a$-node has initially degree $\mbox{$2(n+2)$}$ and its degree increases
by a factor of $\mbox{$n+2$}$ at each generation. Thus, at generation $g$,
the degree of an $a$-node added at generation $g_0$ is,
\begin{equation}
k_a(g-g_0)=2(n+2)^{g-g_0+1} \ \ .
\end{equation}
A $b$-node has initially degree two and its degree increases by a factor
of $m+2$ at each generation. Thus, at generation $g$, the degree of a
$b$-node added at generation $g_0$ is,
\begin{equation}
k_b(g-g_0)=2(m+2)^{g-g_0} \ \ .
\end{equation}
Consequently, the node of degree $k$ at generation $g$ that was added at
generation $g_0$ is given by,
\begin{equation}
g-g_{0a}(k)=\frac{\ln(k/2)}{\ln(n+2)}-1 \ \ ,
\end{equation}
for $a$-nodes and
\begin{equation}
g-g_{0b}(k)=\frac{\ln(k/2)}{\ln(m+2)} \ \ ,
\end{equation}
for $b$-nodes. The degree distribution for the $a$-nodes in the loop is then,
\begin{eqnarray}
P^a_{m,n}(k,g)&=&\frac{\Delta N_a(g_{0a}(k))}{N_a(g)} \nonumber \\
&=&\left[(n+2)(m+2)-1\right]f_{m,n}(g)
\left(\frac{k}{2}\right)^{-\left[1+\frac{\ln(m+2)}{\ln(n+2)}\right]} ,
\end{eqnarray}
where,
\begin{equation}
f_{m,n}(g)=\frac{\left[(n+2)(m+2)\right]^g}{\left[(n+2)(m+2)\right]^g-1} \ \ .
\end{equation}
For the $b$-nodes is,
\begin{eqnarray}
P^b_{m,n}(k,g)&=&\frac{\Delta N_b(g_{0b}(k))}{N_b(g)} \nonumber \\
&=&\frac{(n+2)(m+2)-1}{(n+2)(m+2)} f_{m,n}(g) \left(\frac{k}{2}\right)^{-\left[1+\frac{\ln(n+2)}{\ln(m+2)}\right]} .
\end{eqnarray}
And the total degree distribution $P_{m,n}(k,g)$ is,
\begin{eqnarray}
P_{m,n}(k,g)&=&\frac{\Delta N_a(g_{0a}(k))+\Delta
N_b(g_{0b}(k))}{N(g)} \nonumber \\
&=&f_{m,n}(g) \left[(m+1)\left(\frac{k}{2}\right)^{-\left[1+\frac{\ln(m+2)}{\ln(n+2)}\right]}\right.
\nonumber \\
&
&\left.+\frac{n+1}{n+2}\left(\frac{k}{2}\right)^{-\left[1+\frac{\ln(n+2)}{\ln(m+2)}\right]}
\right] .
\end{eqnarray}
In the limit $g\rightarrow\infty$,
\begin{eqnarray}
P_{m,n}(k)&=&(m+1)\left(\frac{k}{2}\right)^{-\left[1+\frac{\ln(m+2)}{\ln(n+2)}\right]}
\nonumber\\
&
&+\frac{n+1}{n+2}\left(\frac{k}{2}\right)^{-\left[1+\frac{\ln(n+2)}{\ln(m+2)}\right]}
\ \ .
\end{eqnarray}
When $m=n$, $\gamma=2$ is recovered~\cite{Doye05}. For $m\not=n$, the
degree distribution is a sum of two power laws. Asymptotically, for
$k\rightarrow\infty$, $P_{m,n}(k)$ is dominated by the term with
the smallest exponent and thus
\begin{equation}
\gamma=\min\left\{1+\frac{\ln(m+2)}{\ln(n+2)},1+\frac{\ln(n+2)}{\ln(m+2)}
\right\} \ \ .
\end{equation}
The contact network of a bearing is always a scale-free network of
$1<\gamma\leq2$, as shown in Fig.~\ref{fig::gamma}.

Note that the degree exponent is symmetric to permutations of $m$
and $n$. The degree distribution of the entire unit cell is symmetric at
every generation as it is given by,
\begin{eqnarray}
P(k)&=&\frac{(n+1)P_{m,n}(k)+(m+1)P_{n,m}(k)}{m+n+2}
\nonumber \\
&=&\frac{\left[(n+1)(m+1)+\frac{(m+1)^2}{m+2}\right]
       \left(\frac{k}{2}\right)^{-\left[1+\frac{\ln(m+2)}{\ln(n+2)}\right]}}
          {m+n+2} \nonumber \\
&+&\frac{\left[\frac{(n+1)^2}{n+2}+(n+1)(m+1)\right]
         \left(\frac{k}{2}\right)^{-\left[1+\frac{\ln(n+2)}{\ln(m+2)}\right]}}
                {m+n+2} \ \ .
\end{eqnarray}

\subsection{Shortest path and clustering coefficient}
\begin{figure*}
\includegraphics[width=0.485\textwidth]{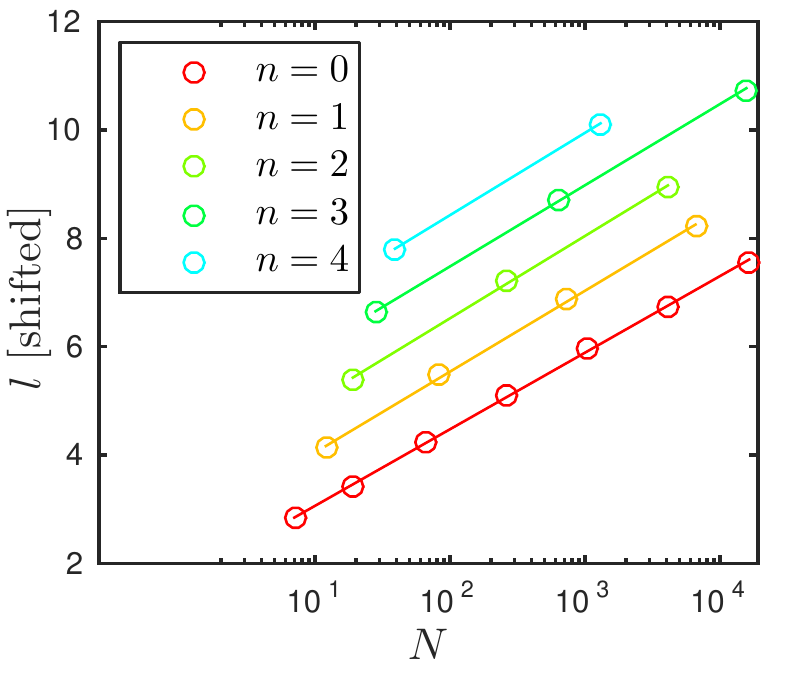}
\includegraphics[width=0.49\textwidth]{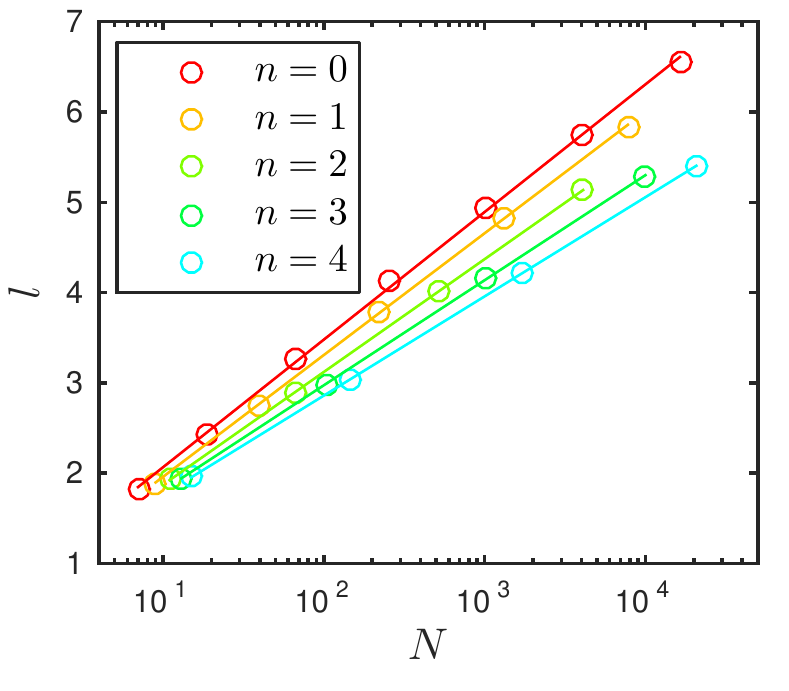}
\caption{\label{fig::slopes}Left: Average shortest path $l$ as a
function of the number of nodes $N$ for $n=m$. Lines were shifted
vertically by adding $n$ for better visualization. Right: Average
shortest path $l$ as a function of the number of sites $N$ for $m=0$.
Lines were also shifted vertically. The lines are just guides to the eye.}
\end{figure*}

We numerically analyze the size dependence of the average shortest path
$l$ for all combinations of $n,m=0,1,2,3,4$. For all cases, we find a
logarithmic scaling of $l$ with the number of nodes, consistent with
small-world networks.  For $m=n$ we find that the prefactor of the
logarithmic scaling is independent on the value of the indices, as also
observed for $\gamma$. If $m\not=n$, then the prefactor changes with $m$
and $n$ as shown in Fig.~\ref{fig::slopes}, for example, for fixed $m=0$,
the prefactor decreases with $n$.

Next, we calculate $C_4$. A new $a$-node has $\mbox{(n+3)(n+2)-1}$ pairs
of neighboring $b$-nodes connected indirectly through one $a$-node. At
each generation, when new neighbors are added to each loop
adjacent to this $a$-node, the number of connected pairs increases by
$\mbox{(n+3)(n+2)/2-1}$.  Hence, an $a$-node that was added to the
network at generation $g_0$, and is part of the network for $t=g-g_0$
generations, has clustering coefficient,
\begin{widetext}
\begin{eqnarray}
C_{4a}(t)&=&\frac{(n+3)(n+2)-1+\left[\frac{(n+3)(n+2)-1}{2}-1\right]\sum_{i=0}^{t-1}k_a(i)}{k_a(t)(k_a(t)-1)/2}
\nonumber \\
&=&\frac{(n+3)(n+2)-1}{(n+2)^{t+1}\left[2(n+2)^{t+1}-1\right]}
+\frac{2(n+2)\left[\frac{(n+3)(n+2)-1}{2}-1\right]\frac{(n+2)^t-1}{n+1}}{(n+2)^{t+1}\left[2(n+2)^{t+1}-1\right]}
\nonumber \\
&\sim&(n+2)^{-t} \text{   as } t\rightarrow \infty \ \ . \nonumber
\end{eqnarray}
\end{widetext}

Initially, for $b$-nodes there is only one pair of neighbors indirectly
connected and $\frac{(m+3)(m+2)}{2}-1$ connections are added per
adjacent loop at each iteration. Thus,
\begin{eqnarray}
C_{4b}(t)&=&\frac{1+\left[\frac{(m+3)(m+2)-1}{2}-1\right]\sum_{i=0}^{t-1}k_b(i)}{k_b(t)(k_b(t)-1)/2}
\nonumber \\
&=&\frac{1+2\left[\frac{(m+3)(m+2)-1}{2}-1\right]\frac{(m+2)^t-1}{m+1}}
{(m+2)^t(2(m+2)^t-1)} \nonumber \\
&\sim&(m+2)^{-t} \text{   as } t\rightarrow \infty. \nonumber
\end{eqnarray}
The argument of the power law is different for $a$- and $b$-nodes as it
depends on $n$ and $m$, respectively. Both $C_{4a}$ and $C_{4b}$
asymptotically vanish. The faster the degree of a node type grows the
faster its $C_4$ falls off.

\subsection{Bond percolation}
We performed simulations of the bond percolation model on a
unit-cell for different pairs of indices $m,n$. As in
Sec.~\ref{sec::spec_percol}, we define the spanning cluster as a set of connected
nodes that include the top and bottom surfaces. Note that the top and bottom
surfaces correspond to different types of nodes, $a$ and $b$, respectively (see
Fig.~\ref{fig::gaskets_stripe}).

For all considered values of $m$ and $n$ we find that the percolation threshold
$p_c$ vanishes in the thermodynamic limit (infinite system size) and the
estimator for the threshold scales as $p_c(N)\sim N^{-\frac{1}{2\nu}}$. Our
results for $m=n$ suggest that $\nu$ does not change with the bearing
indices ($m$ and $n$), like we also found for the degree exponent $\gamma$ in
Sec.~\ref{sec::dec_dist_gen}.  Since the number of nodes in the network grows
exponentially with the generation, we refrain from performing a detailed
size-dependence analysis to obtain $\nu$ with high precision. 

\section{\label{sec::final_remarks}Final remarks}
We studied the contact network of space-filling bearings of loops of size four
in the first family. We proposed a hierarchical rule to construct the network
and provided analytic expressions for the degree distribution and clustering
coefficient. We also studied numerically the shortest path and percolation
properties. We showed that the exponent $\gamma$ changes in the range $(1,2]$
and that is always two when $m=n$. Numerical simulations also suggest that the
correlation exponent $\nu$ for the percolation transition does not change with
the bearing indices. Our networks are bipartite and we find that if $m\not=n$
the degree distribution of the two species scale with different exponents inside
each loop. To our knowledge, this is the first example of an artificial
hierarchical network exhibiting this property which has already been observed
empirically for sexual networks~\cite{Blasio07}.

We are proposing a method to generate deterministic hierarchical scale-free
networks of different $\gamma$ exponents, which are amenable to analytic
treatment. As it was accomplished for the Apollonian network, possible
extensions of our work include the study of their magnetic, spectral, and
dynamical properties~\cite{Moreira06,Oliveira09,Oliveira10}. Other
possibilities include the study of the networks of random space-filling or
three-dimensional bearings, larger loops and second family of bearings.

\begin{acknowledgments}
We acknowledge financial support from the ETH Risk Center, the Brazilian
agencies CNPq, CAPES, FUNCAP, the Brazilian Institute INCT-SC, ERC Advanced
Grant number FP7-319968 of the European Research Council, and the Portuguese
Foundation for Science and Technology (FCT) under the contracts no.
IF/00255/2013 and UID/FIS/00618/2013. JJK thanks ''Studienstiftung des
deutschen Volkes`` for a scholarship.
\end{acknowledgments}
\bibliography{bearingnet.bib}
\end{document}